\documentclass[preprint,prd,nofootinbib,tightenlines,amsmath]{revtex4}
\usepackage{axodraw}
\usepackage{graphics}
\usepackage{epsfig}
\usepackage{bm}% bold math

\begin{document}
\baselineskip=15pt \parskip=5pt

\hspace*{\fill} $\hphantom{-}$

\def\lsim{\mathrel {\vcenter {\baselineskip 0pt \kern 0pt
    \hbox{$<$} \kern 0pt \hbox{$\sim$} }}}
\def\gsim{\mathrel {\vcenter {\baselineskip 0pt \kern 0pt
    \hbox{$>$} \kern 0pt \hbox{$\sim$} }}}

\preprint{}

\title{Rare Decays with a Light $\bm{CP}$-Odd Higgs Boson in the NMSSM}

\author{Xiao-Gang He}
\email{hexg@phys.ntu.edu.tw}
\affiliation{Department of Physics and Center for Theoretical Sciences, National Taiwan University, Taipei,
Taiwan}

\author{Jusak Tandean}
\email{jtandean@yahoo.com}
\affiliation{Department of Physics and Center for Theoretical Sciences,\ National Taiwan University, Taipei,
Taiwan}

\author{G. Valencia}
\email{valencia@iastate.edu}
\affiliation{Department of Physics and Astronomy, Iowa State University, Ames, IA 50011, USA}

\date{\today}

\begin{abstract}

We have previously proposed a light pseudoscalar Higgs boson in the next-to-minimal supersymmetric
standard model (NMSSM), the $A_1^0$, as a candidate to explain the HyperCP observations in
\,$\Sigma^{+}\to p \mu^{+}\mu^{-}$.\,
In this paper we calculate the rates for several other rare decay modes that can help confirm or refute 
this hypothesis.
The first modes we evaluate are  \,$K_L^{}\to\pi\pi A_1^0$,\,  which are interesting because they are
under study by the KTeV Collaboration.
We next turn to  \,$\eta\to\pi\pi A_1^0$,\, which are interesting because they are independent of
the details of the flavor-changing sector of the NMSSM and may be accessible at~DA$\Phi$NE.
For completeness, we also evaluate  \,$\Omega^-\to\Xi^-A_1^0$.\,

\end{abstract}

%\pacs{PACS numbers: }

\maketitle

\section{Introduction}

The HyperCP Collaboration recently observed three events for the rare decay mode
\,$\Sigma^{+}\to p\mu^{+}\mu^{-}$\,  with dimuon invariant masses narrowly clustered around
214.3\,MeV~\cite{Park:2005ek}.
It is possible to account for these events within the standard model (SM) when long-distance
contributions are properly included~\cite{Bergstrom:1987wr,He:2005yn}.
However, the probability that the three events have the same dimuon mass, given the SM
predictions, is less than one percent.
This result has prompted several studies investigating the consequences of a new state with this
mass~\cite{He:2005we,Deshpande:2005mb,Geng:2005ra}.

In particular, it was pointed out that the flavor-changing coupling of the new state to $\bar ds$ has
to be (dominantly) of a pseudoscalar or axial-vector nature to explain why it has not been seen
in  \,$K\to\pi\mu^+\mu^-$.\,  This would still allow  the new particle to be observed
in the other rare modes  \,$K\to\pi\pi\mu^+\mu^-$\,  and  \,$\Omega^-\to\Xi^-\mu^+\mu^-$.\,
Predictions for the new particle's contributing to these modes, consistent with existing constraints,
were made in Refs.~\cite{He:2005we,Deshpande:2005mb}.
These predictions indicate that there could be evidence for the particle in the data already taken by
the KTeV Collaboration, specifically in the mode  \,$K_L\to\pi^0\pi^0\mu^+\mu^-$\, currently being
studied~\cite{priv}.

Beyond the above-mentioned theoretical analyses, to explore the possible consequences of the HyperCP
result in greater detail one has to incorporate some model dependence.
To this end, various ideas have been proposed in the literature~\cite{models,He:2006fr}.
Specifically, we have demonstrated that a~light pseudoscalar Higgs boson in the next-to-minimal
supersymmetric standard model (NMSSM), the~$A_1^0$, could be identified as the possible new particle
responsible for the HyperCP events while satisfying all constraints from kaon and $B$-meson
decays~\cite{He:2006fr}.

Now, it is long known that kaon decays involving a light Higgs boson, such as in the NMSSM, receive
two types of contributions that can be of comparable size: two-quark contributions in which the flavor
change occurs in one-loop processes involving the light $A_1^0$, and four-quark contributions in which
the flavor change occurs via a tree-level standard-model $W$ exchange with the light $A_1^0$
radiated off one of the light quarks~\cite{sdH,Grzadkowski:1992av}.
Not too long ago, we showed that the same situation occurs in the case of light Higgs production in
hyperon decays~\cite{He:2006uu}.

In this paper we revisit the modes  \,$K\to\pi\pi A_1^0$\,  and  \,$\Omega^-\to\Xi^-A_1^0$\, in order
to present a complete prediction within the model suggested in Ref.~\cite{He:2006fr},
plus the possible modifications recently pointed out in Ref.~\cite{Xiangdong:2007vv}.
This differs from the model-independent studies of Refs.~\cite{He:2005we,Deshpande:2005mb} in two
important ways.
Within the NMSSM, we can identify the effective scalar and pseudoscalar couplings of the model-independent
studies with specific one-loop processes.
Here we consider not only the chargino-mediated diagrams of Refs.~\cite{He:2006fr,Hiller:2004ii},
but also the gluino- and neutralino-mediated diagrams discussed in Ref.~\cite{Xiangdong:2007vv}.
In addition, we include the four-quark contributions which are missing in
Refs.~\cite{He:2005we,Deshpande:2005mb}.
These four-quark contributions were shown in Ref.~\cite{He:2006fr} to be essential to evade the bounds
arising from the nonobservation of the $A_1^0$ in  \,$K\to\pi\mu^+\mu^-$\, modes~\cite{kbounds}.
Following our earlier work~\cite{He:2005we,He:2006fr}, we will here assume that
\,${\cal B}(A_1^0 \to \mu^+\mu^-)\sim 100\%$,\,

Additional processes where such a light $A_1^0$ would appear have been recently studied in the literature:
collider signatures for a light $A^0_1$~\cite{Zhu:2006zv}, $B$-meson decays~\cite{Heng:2008rc}, and
radiative quarkonium decays~\cite{Mangano:2007gi}.
The latter are especially useful because, being flavor conserving, they are independent of the specifics
of the one-loop flavor-changing couplings and follow directly from the tree-level couplings of
the $A_1^0$ to down-type quarks.

Finally, in this paper we also consider the modes  \,$\eta\to\pi\pi A_1^0$\, which, like radiative
quarkonium decays, are flavor diagonal and only sensitive to the tree-level couplings of the $A_1^0$.
The prediction for these modes is, therefore, much less model-dependent.
We find a rate two orders of magnitude larger than the corresponding SM rate, that could be probed
at DA$\Phi$NE.

\section{The light $A_1^0$ in the NMSSM}

In this section, we briefly review some features of the NMSSM that are relevant to our study.
The model is an extension of the minimal supersymmetric standard model (MSSM) and provides a~solution
to the so-called $\mu$-problem of the MSSM~\cite{nmssm}.
In the NMSSM, there is a gauge-singlet Higgs field $N$ in addition to the two Higgs fields
$H_u$ and $H_d$ responsible for the  up- and down-type quark masses in the MSSM.
As a result, the physical spectrum of the extended model has two additional neutral Higgs bosons:
one a scalar and the other a pseudoscalar.

We follow the specific model described in Ref.~\cite{Hiller:2004ii}, with suitable modifications.
The superpotential of the model is given by
\begin{eqnarray}
W &=& Q Y_u^{} H_u^{} U + Q Y_d^{} H_d^{} D + L Y_e^{} H_d^{} E + \lambda H_d^{} H_u^{} N
- \mbox{$\frac{1}{3}$} k N^3  \,\,,
\end{eqnarray}
where  $Q$, $U$, $D$, $L$, and $E$ represent the usual quark and lepton fields, $Y_{u,d,e}$
are the Yukawa couplings, and $\lambda$ and $k$ are dimensionless parameters.
The soft-supersymmetry-breaking term in the Higgs potential is
\begin{eqnarray}
V_{\rm soft}^{} &=& m^2_{H_u} |H_u^{}|^2 + m^2_{H_d} |H_d^{}|^2 + m^2_N |N|^2
- \left(\lambda A_\lambda^{}H_d^{}H_u^{}N+\mbox{$\frac{1}{3}$}k A_k^{}N^3+{\rm H.c.}\right)  \,\,,
\end{eqnarray}
and the resulting Higgs potential  has a global $U(1)_R$ symmetry in the limit that the
parameters  \,$A_\lambda,A_k\to0$\,~\cite{Dobrescu:2000yn}.

The NMSSM has two physical $CP$-odd Higgs bosons which are linear combinations of the pseudoscalar
components in $H_{u}$, $H_{d}$, and $N$ in the model mix, with the $A_1^0$ being the lighter
mass-eigenstate with mass given  by
\begin{eqnarray}
m^2_{\cal A} &=& 3 k\, x\, A_k^{}  \,+\,  {\cal O}(1/\tan\beta)
\end{eqnarray}
in the large-$\tan\beta$ limit, where  \,$x=\langle N\rangle$\, is the vacuum expectation value of
$N$  and  $\tan\beta$ is the ratio of vacuum expectation values (VEVs) of the two Higgs doublets.
If the $U(1)_R$ symmetry is broken slightly, the mass of $A^0_1$ becomes naturally small, with
values as low as  \,$\sim$100\,MeV\, phenomenologically
allowed~\cite{Hiller:2004ii,Dobrescu:2000yn,Dobrescu:1999gv}.

In the large-$\tan\beta$ limit, the $A_1^0$ is mostly the singlet pseudoscalar and couples to SM
fields through mixing.
Also in the large-$\tan\beta$ limit, its couplings to fermions are suppressed by a factor of
$\tan\beta$ with respect to those of the $A^0$ in the MSSM~\cite{Hiller:2004ii,Dobrescu:2000yn}.
In particular, this makes the tree-level couplings to up-type quarks negligible.
The tree-level couplings to down-type quarks and charged leptons can be described in terms of one parameter,
\begin{eqnarray}
{\cal L}_{{\cal A}d d}^{}  \,\,=\,\,
-l_d^{} m_d^{}\,\bar d\gamma_5^{}d\,\frac{i A^0_1}{v}  \,\,,  \hspace{2em}
{\cal L}_{{\cal A}\ell}^{}  \,\,=\,\,
-l_d^{} m_\ell^{}\,\bar \ell\gamma_5^{}\ell\, \frac{i A^0_1}{v} \,\,,
\label{44q}
\end{eqnarray}
where the parameter $l_d$ involves both the different Higgs VEVs and
soft-supersymmetry-breaking parameters,
\begin{eqnarray}
l_d^{} \,\,=\,\, \frac{v\,\delta_-^{}}{\sqrt{2}\,x}  \,\,. \hspace{2em}
\end{eqnarray}
with  \,$v=246$\,GeV\, being the electroweak scale and
\,$\delta_-^{}=(A_\lambda-2k x)/(A_\lambda+k x)$.\,
Requiring the mass of the heavier pseudoscalar  not to exceed  \,500\,GeV,\,
Ref.~\cite{Hiller:2004ii} finds a lower bound  \,$|l_d| \gsim 0.1$\, for  \,$\tan\beta=30$.\,
At the same time, the contribution of $A_1^0$ to the muon anomalous magnetic moment results in
an upper bound  \,$|l_d|\lesssim 1.2$~\cite{He:2005we}.

As shown in Ref.~\cite{He:2006fr}, this scenario leads to four-quark contributions that easily
reproduce the HyperCP result. Unfortunately, they are also in conflict with the nonobservation of
the $A_1^0$  in  \,$K\to\pi\mu^+\mu^-$~\cite{kbounds}.
To satisfy these bounds, it is necessary to include contributions from one-loop flavor-changing
diagrams in the NMSSM, which depend in a complicated way on the many parameters of the model.
We can describe them in terms of an effective Lagrangian for the \,$sdA_1^0$\, couplings as
\begin{eqnarray}  \label{LAsd}
{\cal L}_{{\cal A}sd}^{} \,\,=\,\,
\frac{i C_R^{}}{2}\, \bar d(1+\gamma_5^{})s\, A_1^0
\,+\,  \frac{i C_L^{}}{2}\, \bar{d}(1-\gamma_5)s\, A_1^0  \,\,+\,\, {\rm H.c.} \,\,,
\end{eqnarray}
where the parameters  $C_{L,R}^{}$  are generally independent.

In Ref.~\cite{He:2006fr}, we followed Ref.~\cite{Hiller:2004ii} to consider only chargino-mediated
one-loop diagrams in the large-$\tan\beta$ limit.
Furthermore, we selected the supersymmetric parameters so as to suppress the  \,$b\to s$\, transition
and in this way satisfy the nonobservation of $A_1^0$ in  $B$ decay~\cite{bbounds}.
This scenario led to \,$C_L^{}=-C_R^{}m_d^{}/m_s^{}=-2g_{\cal A}^{}\, m_d^{}/v$\,
with  \,$g_{\cal A}^{} \sim 10^{-7}$\,~\cite{He:2006fr}.
More recently, Ref.~\cite{Xiangdong:2007vv} has pointed out a~different scenario in which $C_{L,R}$
also receive contributions from gluino- and neutralino- mediated one-loop diagrams.
Although the gluino-mediated contributions are suppressed by a factor of $\tan\beta$ compared
to the chargino contributions, the former are proportional to the strong coupling $\alpha_s^{}$,
compensating for the suppression factor, and hence can be as important as the latter.
Moreover, in some regions of the parameter space the neutralino-mediated contributions could
be comparable to the gluino-mediated ones~\cite{Xiangdong:2007vv}.
If all the different contributions are similar in size, then $C_{L,R}$ can become effectively independent.

This opens up the possibility of satisfying the kaon bounds without the four-quark contributions by
having \,$C_L\sim-C_R$,\, which results in an effective \,$sdA_1^0$\, coupling that is mostly pseudoscalar.
The HyperCP observation can then be explained as in the model-independent analysis of
Refs.~\cite{He:2005we,Deshpande:2005mb}.
However, to have  \,$C_L\sim-C_R$\, requires some sort of fine tuning.
Furthermore, the four-quark contributions may not necessarily be negligible.
In our analysis, we will thus keep $C_L$ and $C_R$ independent and constrain them with data.
Also, we will assume that $CP$ is conserved and hence $C_{L,R}$ are real.

\section{$\bm{|\Delta S|=1}$\,  decays\label{ds}}

\subsection{Two-quark contributions\label{2q}}

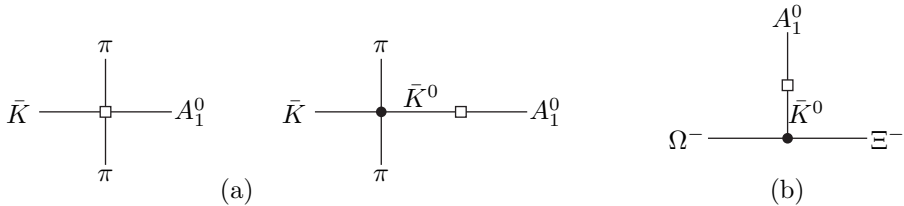
\begin{figure}[bh]
\begin{picture}(90,90)(-45,-40)
\Text(-33,0)[]{\footnotesize$\bar K$} \Line(-25,0)(25,0)
\Line(0,-20)(0,20) \Text(0,25)[]{\footnotesize$\pi$} \Text(0,-24)[]{\footnotesize$\pi$}
\Text(32,0)[]{\footnotesize$A_1^0$} \BBoxc(0,0)(4,4)
\end{picture}
\begin{picture}(2,90)(-1,-40) \Text(0,-30)[]{\footnotesize(a)} \end{picture}
\begin{picture}(120,90)(-50,-40)
\Text(-33,0)[]{\footnotesize$\bar K$} \Line(-25,0)(55,0)
\Line(0,-20)(0,20) \Text(0,25)[]{\footnotesize$\pi$} \Text(0,-24)[]{\footnotesize$\pi$}
\Text(15,6)[]{\footnotesize$\bar K^0$} \Text(62,0)[]{\footnotesize$A_1^0$}
\BBoxc(30,0)(4,4) \Vertex(0,0){2}
\end{picture}
\begin{picture}(150,90)(-80,-30)
\Text(-38,0)[]{\footnotesize$\Omega^-$} \Line(-30,0)(30,0) \Line(0,0)(0,40)
\Text(7,10)[]{\footnotesize$\bar{K}^0$} \Text(0,45)[]{\footnotesize$A_1^0$}
\Text(38,0)[]{\footnotesize$\Xi^-$} \BBoxc(0,20)(4,4) \Vertex(0,0){2}
\Text(0,-20)[]{\footnotesize(b)}
\end{picture}
\vspace{-2ex}
\caption{\label{2qdiag}%
Diagrams contributing to (a) $\bar K\to\pi\pi A_1^0$\, and (b) $\Omega^-\to\Xi^-A_1^0$
arising from  ${\cal L}_{{\cal A}sd}^{}$  at leading order in  $\chi$PT.
The square vertices come from  ${\cal L}_{\cal A}^{}$  in  Eq.~(\ref{LA}),
and the solid dots from  ${\cal L}_{\rm s}^{}$  in  Eq.~(\ref{Ls}).
}
\end{figure}

To evaluate hadronic amplitudes induced by the $sd A_1^0$ interactions, we employ chiral perturbation
theory ($\chi$PT).
Thus, the leading-order chiral realization of ${\cal L}_{{\cal A}sd}$ above is ${\cal L}_{\cal A}$ in
Eq.~(\ref{LA}) in Appendix~\ref{lags}, which also contains other relevant chiral Lagrangians.
From  ${\cal L}_{\cal A}$  and the chiral strong Lagrangian ${\cal L}_{\rm s}$ in Eq.~(\ref{Ls}),
we derive the leading-order diagrams shown in Fig.~\ref{2qdiag}  for  \,$\bar K\to\pi\pi A_1^0$\,
and  \,$\Omega^-\to\Xi^-A_1^0$.\,
The resulting amplitudes are
\begin{eqnarray}
{\cal M}_{2q}^{}\bigl(\bar{K}^0\to\pi^+\pi^-A_1^0\bigr) \,\,=\,\,
\frac{B_0^{}\, \bigl(C_L^{}-C_R^{}\bigr)}{\sqrt8\, f}\,\,
\frac{m_{\cal A\pi^+}^2-m_\pi^2-m_{\cal A}^2}{m_K^2-m_{\cal A}^2}  \,\,,
\end{eqnarray}
\begin{eqnarray}
{\cal M}_{2q}^{}\bigl(\bar K^0\to\pi^0\pi^0A_1^0\bigr)  \,\,=\,\,
\frac{B_0^{}\, \bigl(C_L^{}-C_R^{}\bigr)}{4\sqrt2\,f}\,\,
\frac{m_K^2-m_{\cal A}^2-m_{\pi^0\pi^0}^2}{m_K^2-m_{\cal A}^2} \,\,,
\end{eqnarray}
\begin{eqnarray}
{\cal M}_{2q}^{}\bigl(\Omega^-\to\Xi^-A_1^0\bigr)  \,\,=\,\,
\frac{i B_0^{}\,{\cal C}}{2}\,\, \frac{C_R^{}-C_L^{}}{m_K^2-m_{\cal A}^2}\,\,
(p_{\cal A}^{})_\mu^{}\, \bar u_\Xi^{}u_\Omega^\mu  \,\,,
\end{eqnarray}
where  \,$m_{XY}^2=(p_X^{}+p_Y^{})^2$.\,
The same Lagrangians also yield
% 2q
\begin{eqnarray}   \label{M2q(K->piA)}
{\cal M}_{2q}^{}\bigl(K^+\to\pi^+A_1^0\bigr)  \,\,=\,\,
-\sqrt2\, {\cal M}_{2q}^{}\bigl(K^0\to\pi^0A_1^0\bigr)  \,\,=\,\,
\frac{i B_0^{}}{2} \bigl(C_L^*+C_R^*\bigr) \,\,,
\end{eqnarray}
\begin{eqnarray}   \label{M2q(S->pA)}
{\cal M}_{2q}^{}\bigl(\Sigma^+\to p A_1^0\bigr) &=&
i\bigl(C_L^{}+C_R^{}\bigr)\frac{B_0^{}}{2}\,\,
\frac{m_\Sigma^{}-m_N^{}}{m_K^2-m_\pi^2}\,\, \bar p\Sigma^+
\nonumber \\ && -\,\,
i\bigl(C_R^{}-C_L^{}\bigr)(D-F)\frac{B_0^{}}{2}\,\,
\frac{m_\Sigma^{}+m_N^{}}{m_K^2-m_{\cal A}^2}\,\, \bar p\gamma_5^{}\Sigma^+ \,\,,
\end{eqnarray}
previously derived in Refs.~\cite{He:2006uu,Xiangdong:2007vv}.
Hence we also adopt  \,$D-F=0.25$.\,

\subsection{Four-quark contributions\label{4q}}

{From}  ${\cal L}_{\rm s,w}^{({\cal A})}$ given in Appendix~\ref{lags}, we obtain the leading-order
diagrams shown in Figs.~\ref{4qdiag} and~\ref{4qdiag'}  for  \,$\bar K\to\pi\pi A_1^0$\, and
\,$\Omega^-\to\Xi^-A_1^0$.\,
The resulting amplitudes are
\begin{eqnarray}
{\cal M}_{4q}^{}\bigl(\bar K^0\to\pi^+\pi^-A_1^0\bigr) \,\,=\,\,
\sum_{i=1}^8 {\cal M}_i^{+-}  \,\,,
\end{eqnarray}
\begin{eqnarray}
{\cal M}_{4q}^{}\bigl(\bar K^0\to\pi^0\pi^0A_1^0\bigr) \,\,=\,\,
\sum_{i=1}^8 {\cal M}_i^{00}  \,\,,
\end{eqnarray}
\begin{eqnarray}
{\cal M}_{4q}^{}\bigl(\Omega^-\to\Xi^-A_1^0\bigr)  \,\,=\,\,
\frac{i B_{\Xi^-\pi^0}^{}\, f}{2\, v}
\bigl(-b_\pi^{}+b_\eta^{}\, c_\theta^{}+b_{\eta'}^{}\, s_\theta^{}\bigr)\, l_d^{}\,
(p_{\cal A}^{})_\mu^{}\, \bar{u}_\Xi^{} u_\Omega^\mu  \,\,,
\end{eqnarray}
where the expressions for  ${\cal M}_i^{+-,00}$  and  $b_{\pi,\eta,\eta'}$  have been collected
in Appendix~\ref{Mi}, and $B_{\Xi^-\pi^0}$ is related in $\chi$PT to
the dominant $P$-wave amplitude  for  \,$\Omega^-\to\Xi^-\pi^0$\,  by
\,${\cal M}(\Omega^-\to\Xi^-\pi^0)=
i B_{\Xi^-\pi^0}\,(p_\pi^{})_\mu^{}\, \bar u_\Xi^{}u_\Omega^\mu$.\,
Hence the \,$\Omega^-\to\Xi^-\pi^0$\,  data yields  \,$B_{\Xi^-\pi^0}=-8.17\times10^{-7}$.\,
We note that the $\tilde{\gamma}_8^{}$ contributions to
${\cal M}_{4q}^{}\bigl(\bar K\to\pi\pi A_1^0\bigr)$ cancel completely, which is expected due to
the fact that the $\tilde{\gamma}_8^{}$ terms in ${\cal L}_{\rm w}^{({\cal A})}$ could be rotated
away if the baryonic part were absent~\cite{Grzadkowski:1992av}.
We also note that the $\tilde{\gamma}_8^{}$ contribution to \,$\Omega^-\to\Xi^-\pi^0$\, appears only
at next-to-leading order.

\begin{figure}[b]
\begin{picture}(80,90)(-40,-50)
\Text(-26,0)[]{\footnotesize$\bar K$} \Line(-20,0)(20,0)
\Line(0,-20)(0,20) \Text(0,24)[]{\footnotesize$\pi$} \Text(0,-24)[]{\footnotesize$\pi$}
\Text(26,0)[]{\footnotesize$A_1^0$} \CBoxc(0,0)(4,4){Black}{Black}
\end{picture}
\begin{picture}(110,90)(-40,-50)
\Text(-26,0)[]{\footnotesize$\bar K$} \Line(-20,0)(50,0) \Vertex(0,0){2}
\Line(0,-20)(0,20) \Text(0,24)[]{\footnotesize$\pi$} \Text(0,-24)[]{\footnotesize$\pi$}
\Text(15,5)[]{\footnotesize$\bar K$} \Text(56,0)[]{\footnotesize$A_1^0$} \CBoxc(30,0)(4,4){Black}{Black}
\end{picture}
\begin{picture}(80,90)(-40,-50)
\Text(-26,0)[]{\footnotesize$\bar K$} \Line(-20,0)(20,0) \Vertex(0,0){2} \Vertex(0,-20){2}
\Line(0,-40)(0,20) \Text(0,24)[]{\footnotesize$\pi$} \Text(0,-44)[]{\footnotesize$\pi$}
\Text(5,-10)[]{\footnotesize$\bar K$} \Text(26,0)[]{\footnotesize$A_1^0$} \CBoxc(0,-20)(4,4){Black}{Black}
\end{picture}
\begin{picture}(110,90)(-40,-50)
\Text(-26,0)[]{\footnotesize$\bar K$} \Line(-20,0)(50,0) \Text(15,5)[]{\footnotesize$\cal P$}
\Line(0,-40)(0,20) \Text(0,24)[]{\footnotesize$\pi$} \Text(0,-44)[]{\footnotesize$\pi$}
\Text(5,-10)[]{\footnotesize$\bar K$} \Text(56,0)[]{\footnotesize$A_1^0$}
\CBoxc(0,-20)(4,4){Black}{Black} \Vertex(0,0){2} \Vertex(0,-20){2} \Vertex(30,0){2}
\end{picture}
\begin{picture}(110,90)(-40,-50)
\Text(-26,0)[]{\footnotesize$\bar K$} \Line(-20,0)(50,0) \Text(15,5)[]{\footnotesize$\cal P$}
\Line(30,-20)(30,20) \Text(30,24)[]{\footnotesize$\pi$} \Text(30,-24)[]{\footnotesize$\pi$}
\Text(56,0)[]{\footnotesize$A_1^0$} \CBoxc(0,0)(4,4){Black}{Black} \Vertex(30,0){2}
\end{picture}
\\
\begin{picture}(120,70)(-40,-35)
\Text(-26,0)[]{\footnotesize$\bar K$} \Line(-20,0)(50,0) \Line(0,-20)(0,20)
\Text(0,24)[]{\footnotesize$\pi$} \Text(0,-24)[]{\footnotesize$\pi$} \Text(15,5)[]{\footnotesize$\cal P$}
\Text(56,0)[]{\footnotesize$A_1^0$} \CBoxc(0,0)(4,4){Black}{Black} \Vertex(30,0){2}
\end{picture}
\begin{picture}(150,70)(-40,-35)
\Text(-26,0)[]{\footnotesize$\bar K$} \Line(-20,0)(80,0) \Text(15,5)[]{\footnotesize$\bar K$}
\Line(0,-20)(0,20) \Text(0,24)[]{\footnotesize$\pi$} \Text(0,-24)[]{\footnotesize$\pi$}
\Text(45,5)[]{\footnotesize$\cal P$} \Vertex(0,0){2}
\Text(86,0)[]{\footnotesize$A_1^0$} \CBoxc(30,0)(4,4){Black}{Black} \Vertex(60,0){2}
\end{picture}
\begin{picture}(150,70)(-40,-35)
\Text(-26,0)[]{\footnotesize$\bar K$} \Line(-20,0)(80,0)
\Line(30,-20)(30,20) \Text(30,24)[]{\footnotesize$\pi$} \Text(30,-24)[]{\footnotesize$\pi$}
\Text(15,5)[]{\footnotesize$\cal P$} \Text(45,5)[]{\footnotesize$\cal P$} \Vertex(30,0){2}
\Text(86,0)[]{\footnotesize$A_1^0$} \CBoxc(0,0)(4,4){Black}{Black} \Vertex(60,0){2}
\end{picture}
\vspace{-2ex}
\caption{\label{4qdiag}%
Diagrams contributing to  $\,\bar K\to\pi\pi A_1^0$\,  arising from four-quark operators,
where  \,${\cal P}=\pi^0,\eta,\eta'$.\,
The dots come from  ${\cal L}_{\rm s}^{({\cal A})}$  in  Eqs.~(\ref{Ls}) and~(\ref{LsP}),
whereas the square vertices are from  ${\cal L}_{\rm w}^{({\cal A})}$  in  Eqs.~(\ref{Lw})
and~(\ref{LwP}).
}
\end{figure}
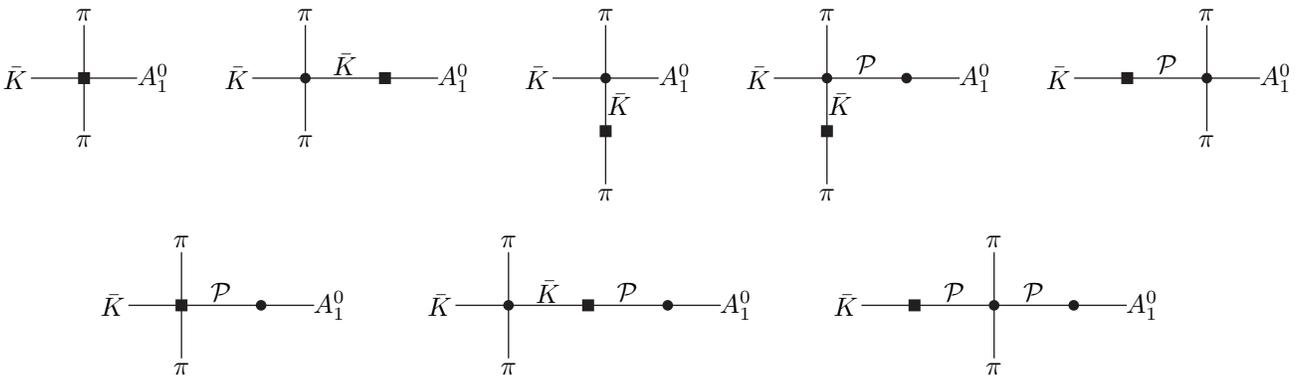
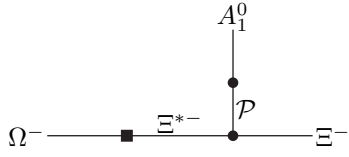
\begin{figure}[tb]
\begin{picture}(110,80)(-40,-15)
\Text(-38,0)[]{\footnotesize$\Omega^-$} \Line(-30,0)(70,0)
\Line(40,0)(40,40) \Text(45,10)[]{\footnotesize$\cal P$}
\Text(40,45)[]{\footnotesize$A_1^0$} \Text(20,5)[]{\footnotesize$\Xi^{*-}$}
\Text(78,0)[]{\footnotesize$\Xi^-$} \CBoxc(0,0)(4,4){Black}{Black} \Vertex(40,0){2} \Vertex(40,20){2}
\end{picture}
\vspace{-2ex}
\caption{\label{4qdiag'}%
Diagram contributing to  $\,\Omega^-\to\Xi^-A_1^0$\,  arising from four-quark operators.
}
\end{figure}

For  \,$K\to\pi A_1^0$\, and  \,$\Sigma^+\to p A_1^0$,\,  the four-quark amplitudes were previously
calculated in Ref.~\cite{He:2006uu}.
For  \,$l_u^{}=0$,\,  they can be rewritten as\footnote{
There is a typo in the last line of Eq. (70) in Ref.\cite{He:2006uu}.
The term \,$-(l_d^{}+l_u^{})m_\pi^2$\, should be corrected to
\,$-(3l_d^{}+l_u^{})m_\pi^2$.\,
This error, however, did not occur in our computation.}
% 4q
\begin{eqnarray}   \label{M4q(K+->piA)}
{\cal M}_{4q}^{}\bigl(K^+\to\pi^+A_1^0\bigr)  &=&
\frac{i}{6 v} \Bigl[
3b_\pi^{}\, \bigl(m_{\cal A}^2-m_\pi^2\bigr) \,+\,
\bigl(b_\eta^{}c_\theta^{}+b_{\eta'}^{}s_\theta^{}\bigr)\bigl(2 m_K^2+m_\pi^2-3 m_{\cal A}^2\bigr)
\nonumber \\ && \hspace*{4ex}
-\,\, \sqrt8\, \bigl(b_\eta^{}s_\theta^{}-b_{\eta'}^{}c_\theta^{}\bigr)\bigl(m_K^2-m_\pi^2\bigr)
\Bigr] \gamma_8^*\, l_d^{} \,\,,
\end{eqnarray}
\begin{eqnarray}   \label{M4q(K0->piA)}
{\cal M}_{4q}^{}\bigl(K^0\to\pi^0A_1^0\bigr)  &=&
\frac{i\sqrt2}{12 v} \Bigl[
3b_\pi^{}\, \bigl(2 m_K^2-m_\pi^2-m_{\cal A}^2\bigr) \,-\,
\bigl(b_\eta^{}c_\theta^{}+b_{\eta'}^{}s_\theta^{}\bigr)\bigl(2 m_K^2+m_\pi^2-3 m_{\cal A}^2\bigr)
\nonumber \\ && \hspace*{6ex}
+\,\,
\sqrt8\, \bigl(b_\eta^{}s_\theta^{}-b_{\eta'}^{}c_\theta^{}\bigr)\bigl(m_K^2-m_\pi^2\bigr)
\Bigr] \gamma_8^*\, l_d^{} \,\,,
\end{eqnarray}
\begin{eqnarray}   \label{M4q(S->pA)}
{\cal M}_{4q}^{}(\Sigma^+\to p {\cal A})  \,\,=\,\,
\frac{f\, l_d^{}}{2 v} \bigl(-b_\pi^{}\,+\,b_\eta^{}c_\theta^{}\,+\,b_{\eta'}^{}s_\theta^{}\bigr) \,\,
i \bar p\, \bigl(A_{p\pi^0}^{}-B_{p\pi^0}^{}\gamma_5^{}\bigr) \,\Sigma^+
\end{eqnarray}
where  \,$A_{p\pi^0}=-3.25\times10^{-7}$\,  and
\,$B_{p\pi^0}=26.67\times10^{-7}$,\,  up to an overall sign,  extracted from
\,$\Sigma^+\to p\pi^0$\,  data.

\subsection{Total contributions\label{tot}}

\begin{figure}[htb]
\vspace{1ex}
\includegraphics[width=3in]{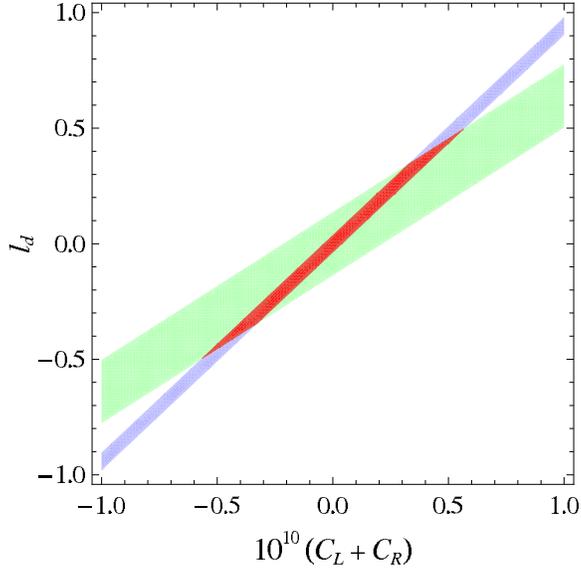}   \vspace{-1ex}
\caption{\label{bounds(K->piA)}Regions in the \,$(C_L$$+$$C_R,l_d)$\, parameter
space allowed by  \,$K^+\to\pi^+\mu^+\mu^-$ (shaded, blue) and
\,$K_S^{}\to\pi^0\mu^+\mu^-$ (lightly shaded, green).
The overlap (dark, red) band covers points that satisfy both constraints.
}
\end{figure}

In this section we present numerical results for the different modes including all contributions to
the respective amplitudes.
We begin by determining the region in the \,$(C_L$$+$$C_R,l_d)$\, parameter space that is allowed by
both the  \,$K^+\to\pi^+\mu^+\mu^-$\, and \,$K_S\to\pi^0\mu^+\mu^-$\, constraints.
We show this in Fig.~\ref{bounds(K->piA)}.
Notice that only small values of \,$C_L$$+$$C_R$\, are allowed.
This corresponds to the conclusion of the analyses of Refs.~\cite{He:2005we,Deshpande:2005mb}  that
the effective $sd A_1^0$ scalar coupling  is severely constrained by these decay modes.
That case, without the four-quark contributions, corresponds to \,$l_d=0$\, in this plot.
The inclusion of the four-quark contributions does not change this conclusion, but simply shifts
the allowed region due to the interplay between the two- and four-quark contributions.

\begin{figure}[tb]
\vspace{1ex}
\includegraphics[width=3in]{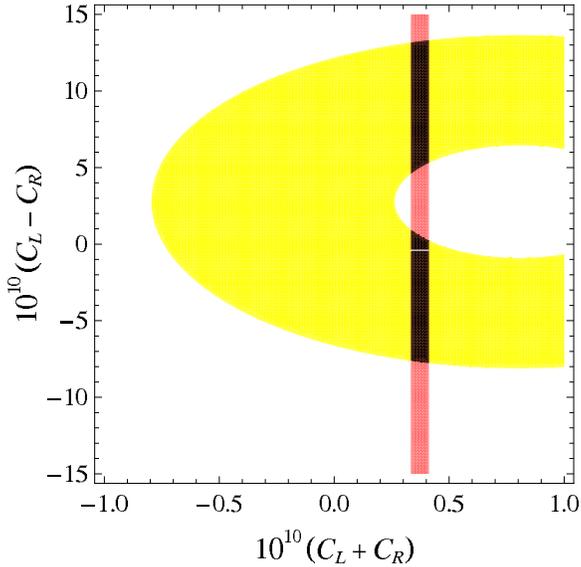}   \vspace{-1ex}
\caption{\label{hypercp}Regions in the \,$(C_L$$+$$C_R,C_L$$-$$C_R)$\,
parameter space reproducing the HyperCP result for  \,$\Sigma^+\to p\mu^+\mu^-$\, (lightly shaded, yellow)
and respecting the  \,$K\to\pi\mu^+\mu^-$\,  bounds (darkly shaded, red) for  \,$l_d=0.35$.\,
The overlap (black) areas cover points satisfying both the hyperon and kaon constraints.
The unshaded (white) region on the vertical band corresponds to the special case discussed in
Ref.~\cite{He:2006fr}.}
\end{figure}

For definiteness, we select \,$l_d^{}=0.35$\, as in Ref.~\cite{He:2006fr} and study the allowed region
in the \,$(C_L$$+$$C_R,C_L$$-$$C_R)$\, parameter space.
We display in Fig.~\ref{hypercp} the lightly shaded (yellow) region that reproduces the HyperCP result for
\,$\Sigma^+\to p\mu^+\mu^-$\, (at the one-sigma level combining statistical and systematic errors in quadrature).
The darkly shaded (red) vertical band covers the region that satisfies the constraints from
the nonobservation  of $A_1^0$ in  \,$K\to\pi\mu^+\mu^-$\, modes.
For  \,$l_d^{}=0$,\, these (yellow and red) areas would both be centered at the origin.
The (black) overlap between these regions is the allowed parameter space that we use for our predictions.
Also displayed on the vertical band is an unshaded (white) thin area corresponding to the
\,$C_L^{}=-C_R^{}m_d^{}/m_s^{}=-2g_{\cal A}^{}\, m_d^{}/v$\,  scenario of Ref.~\cite{He:2006fr}.

\begin{figure}[tb]
\vspace{1ex}
\includegraphics[width=5in]{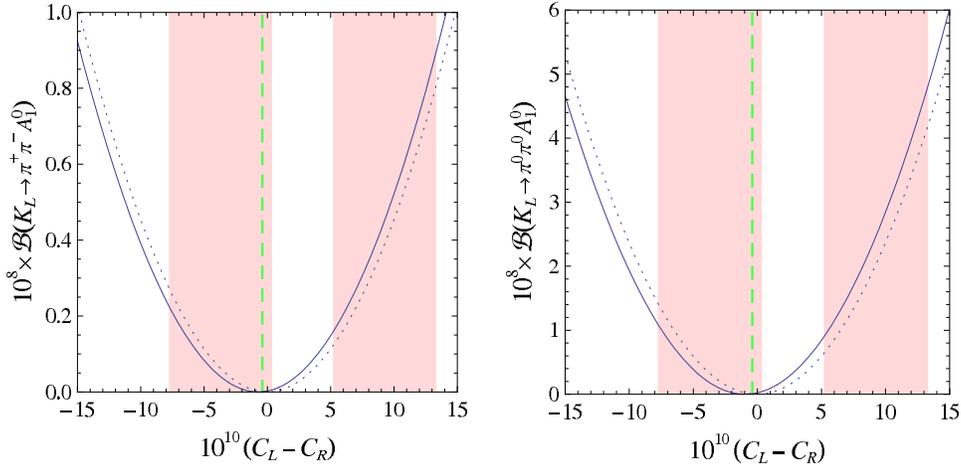} \vspace{-1ex}
\caption{\label{predictions}Predicted branching ratios (solid curves) for  \,$K_L^{}\to\pi^+\pi^-A_1^0$\,
and  \,$K_L^{}\to\pi^0\pi^0A_1^0$\, with \,$l_d^{}=0.35$\, as functions of \,$C_L^{}$$-$$C_R^{}$.
The dotted curves result from the two-quark contributions alone.
The shaded (pink) bands indicate the allowed ranges of \,$C_L^{}$$-$$C_R^{}$\, as determined from
Fig.~\ref{hypercp}.
Each vertical (green) dashed line corresponds to the special case discussed in
Ref.~\cite{He:2006fr}.}
\end{figure}

With these results, we show in Fig.~\ref{predictions} the predicted branching
ratios (solid curves) for \,$K_L^{}\to\pi^+\pi^-A_1^0$\, and \,$K_L^{}\to\pi^0\pi^0A_1^0$\, as
functions of  \,$C_L^{}$$-$$C_R^{}$\, for \,$l_d^{}=0.35$\, and \,$C_L^{}+C_R^{}=4\times 10^{-11}$.\,
The range of each of these predictions over the allowed values of \,$C_L^{}$$-$$C_R^{}$\, is larger than
that obtained in Ref.~\cite{He:2005we}, due partly to the presence of the four-quark contributions
and partly to the uncertainty in the HyperCP measurement.
Each of the solid curves has a minimum that is not zero, as the two- and four-quark contributions
have different kinematical dependences and hence do not cancel in general.
The rates for most of the allowed regions are significantly large, but those around the minima
may be too small to be observed.
For comparison, we also show in Fig.~\ref{hypercp} dotted curves representing the branching ratios
obtained from the two-quark contributions alone and vertical (green) dashed lines each indicating
the narrow range of  \,$C_L^{}$$-$$C_R^{}$\, found in the scenario of Ref.~\cite{He:2006fr}.

\begin{figure}[b]
\vspace{1ex}
\includegraphics[width=3in]{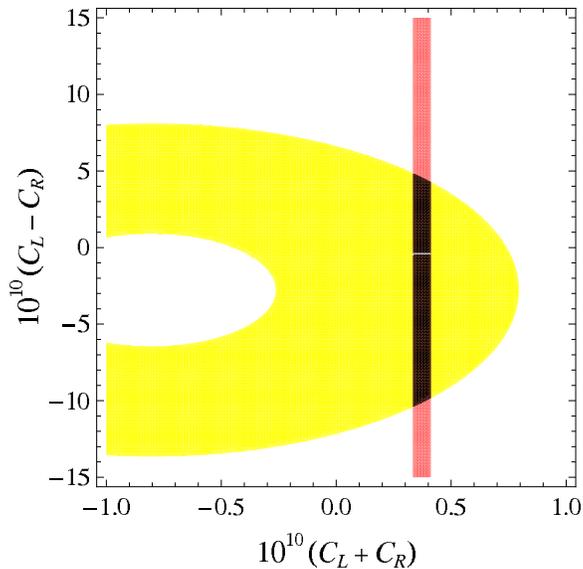}   \vspace{-1ex}
\caption{\label{hypercp'}The same as Fig.~\ref{hypercp}, except that the relative sign between
the two- and four-quark contributions to \,$\Sigma^+\to p\mu^+\mu^-$\, is the opposite.
}
\end{figure}

\begin{figure}[tb]
\vspace{1ex}
\includegraphics[width=5in]{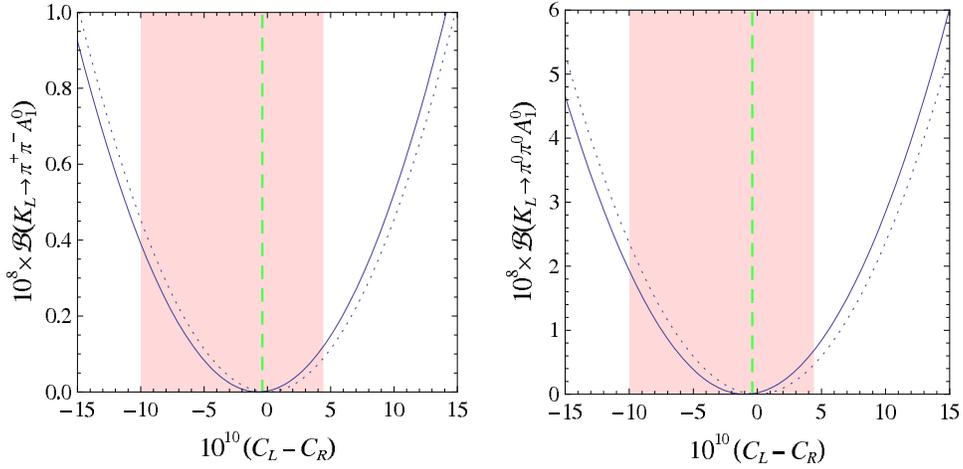} \vspace{-1ex}
\caption{\label{predictions'}The same as Fig.~\ref{predictions}, except that the allowed range of
\,$C_L^{}$$-$$C_R^{}$\, is from Fig.~\ref{hypercp'}.
}
\end{figure}

Since the values of $A_{p\pi^0}$ and $B_{p\pi^0}$ in Eq.~(\ref{M4q(S->pA)}) are determined from
experiment only up to an overall sign, we should also consider the possibility that the two and
four-quark contributions to \,$\Sigma^+\to p\mu^+\mu^-$\, have a different relative sign.
This yields a different allowed range of \,$C_L$$-$$C_R$,\, as can be seen in Fig.~\ref{hypercp'}.
We display the resulting predictions for \,$K_L^{}\to\pi\pi A_1^0$\, in Fig.~\ref{predictions'},
whose ranges over the allowed regions turn out to be roughly only half as large as those in
Fig.~\ref{predictions}, respectively.


\begin{figure}[b]
\vspace{1ex}
\includegraphics[width=3in]{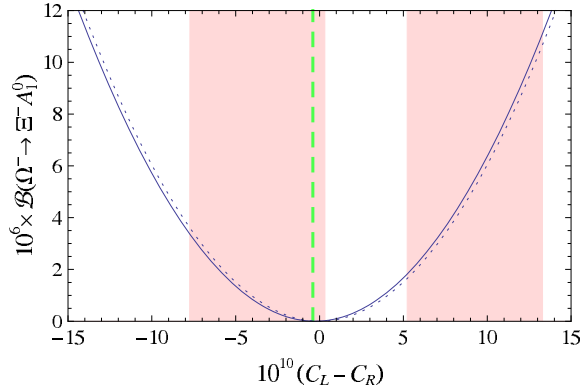} \vspace{-1ex}
\caption{\label{omega}Predicted branching ratio (solid curve) for  \,$\Omega^-\to\Xi^-A_1^0$\,
with \,$l_d^{}=0.35$\, as function of \,$C_L^{}$$-$$C_R^{}$.
The dotted curve results from the two-quark contributions alone.
The shaded (pink) bands indicate the allowed ranges of \,$C_L^{}$$-$$C_R^{}$\, as determined from
Fig.~\ref{hypercp}.
The vertical (green) dashed line corresponds to the special case of Ref.~\cite{He:2006fr}.}
\end{figure}

\begin{figure}[tb]
\vspace{1ex}
\includegraphics[width=3in]{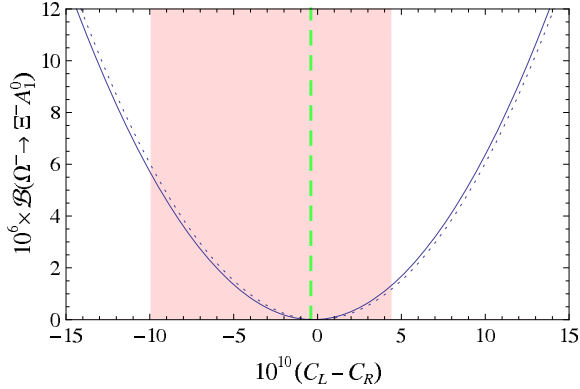} \vspace{-1ex}
\caption{\label{omega'}The same as Fig.~\ref{omega}, except that the allowed range of
\,$C_L^{}$$-$$C_R^{}$\, is from Fig.~\ref{hypercp'}.}
\end{figure}

Finally, from the results of Fig.~\ref{hypercp} we display the predicted branching ratio (solid curve)
for  \,$\Omega^-\to\Xi^-A_1^0$\, in Fig.~\ref{omega}.
The range of the prediction over the allowed values of \,$C_L$$-$$C_R$\, is again larger than
that obtained in Ref.~\cite{He:2005we} due to the presence of the four-quark contributions
as well as to the experimental error.
The best limit for this mode currently available was reported by HyperCP~\cite{solomey},
\,${\cal B}(\Omega^-\to\Xi^-\mu^+\mu^-)<6.1\times 10^{-6}$\, at  90\% C.L.,
whereas the standard-model prediction is
\,${\cal B}_{\rm SM}(\Omega^-\to\Xi^-\mu^+\mu^-)=6.6\times10^{-8}$~\cite{Safadi:1987qq}.
Therefore, the \,$\Omega^-\to\Xi^-A_1^0$\, rate for most of the allowed regions is substantial, 
but the curve has a zero, around which the rate is too small to be observed.
The significant enhancement possible with respect to the SM rate lends support to
pursuing a future $\Omega^-$ experiment~\cite{Kaplan:2007nn}.
For comparison, we also display in Fig.~\ref{omega} the dotted curve representing the branching
ratio obtained from only the two-quark contributions and the vertical (green) dashed lines
corresponding to the special case of Ref.~\cite{He:2006fr}.
In Fig.~\ref{omega'} we show the corresponding prediction with the allowed range of
\,$C_L$$-$$C_R$\, from Fig.~\ref{hypercp'}.


\section{Flavor-conserving decays \,\bm{$\eta\to\pi\pi A_1^0$}}

These modes are special because they involve only flavor-diagonal interactions.
As such, they are not sensitive to the unknown parameters in the flavor sector of the model that
give rise to the two-quark amplitudes.
The predicted rates follow only from the tree-level diagonal couplings of $A_1^0$ and in this way
they are similar to the radiative quarkonium decays proposed in Ref.~\cite{Mangano:2007gi}.
These $\eta$ decays are also analogous to the $\eta$ decay with a light $CP$-even Higgs boson
which was severely constrained by data~\cite{Prades:1990vn}.

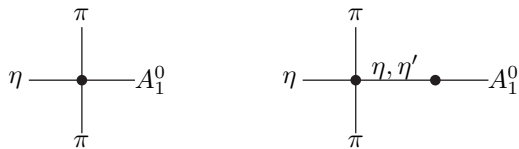
\begin{figure}[b]
\begin{picture}(100,70)(-50,-30)
\Text(-25,0)[]{\footnotesize$\eta$} \Line(-20,0)(20,0)
\Line(0,-20)(0,20) \Text(0,25)[]{\footnotesize$\pi$} \Text(0,-23)[]{\footnotesize$\pi$}
\Text(26,0)[]{\footnotesize$A_1^0$} \Vertex(0,0){2}
\end{picture}
\begin{picture}(140,70)(-50,-30)
\Text(-25,0)[]{\footnotesize$\eta$} \Line(-20,0)(50,0) \Line(0,-20)(0,20)
\Text(0,25)[]{\footnotesize$\pi$} \Text(0,-23)[]{\footnotesize$\pi$}
\Text(15,5)[]{\footnotesize$\eta,\eta'$}
\Text(56,0)[]{\footnotesize$A_1^0$} \Vertex(0,0){2} \Vertex(30,0){2}
\end{picture}
\vspace{-2ex}
\caption{\label{etadiag}%
Diagrams contributing to  \,$\eta\to\pi\pi A_1^0$\,  induced by flavor-diagonal couplings
of  $A_1^0$  to light quarks.
The dots come from  ${\cal L}_{\rm s}^{({\cal A})}$  in  Eqs.~(\ref{Ls}) and~(\ref{LsP}),
}
\end{figure}

The leading-order amplitude for \,$\eta\to\pi\pi A_1^0$\, comes from the two diagrams in Fig.~\ref{etadiag}.
It is the same for  \,$\eta\to\pi^+\pi^-A_1^0$\,   and  \,$\eta\to\pi^0\pi^0 A_1^0$,\,
\begin{eqnarray}
{\cal M}\bigl(\eta\to\pi \pi A_1^0\bigr)  &=&
\frac{\sqrt3\, m_\pi^2}{18 f v} \Bigl[ 3 \bigl(c_\theta^{}-\sqrt2\,s_\theta^{}\bigr)
\,+\, b_\eta^{}\,\bigl(1-\sqrt8\,c_\theta^{}s_\theta^{}+s_\theta^2\bigr)
\,+\,
b_{\eta'}^{}\, \bigl(\sqrt2-c_\theta^{}s_\theta^{}-\sqrt8\,s_\theta^2\bigr) \Bigr] l_d^{} \,\,.
\nonumber \\
\end{eqnarray}

For an  $\eta$-$\eta^\prime$ mixing angle of  \,$\theta=-19.7^\circ$,\,  we then find
\begin{eqnarray}   \label{etawitha}
{\cal B}\bigl(\eta\to\pi^+\pi^-A_1^0\bigr) &=& 5.4\times 10^{-7}\, l_d^2 \,\,, \vphantom{\big|}
\\
{\cal B}\bigl(\eta\to\pi^0\pi^0A_1^0\bigr) &=& 3.2\times 10^{-7}\, l_d^2 \,\,. \vphantom{\Big|}
\end{eqnarray}
Allowing the mixing angle to vary between $-15^\circ$ and $-25^\circ$ would result in $20\%$ changes.
The rate for the neutral-pion mode is not exactly half the rate for the charged-pion mode because
we have used physical masses for the numerical estimate.

The best limit currently available on any of these modes comes from the CELCIUS/WASA collaboration.
At the 90\% C.L. they find~\cite{Berlowski:2008zz}.
\begin{equation}
{\cal B}(\eta \to \pi^+ \pi^- \mu^+\mu^-) \,\,<\,\, 3.6 \times 10^{-4}  \,\,.
\end{equation}
Presently this does not place a stringent bound on the coupling $l_d$, giving  \,$|l_d|<26$.\,
Nevertheless, Eq.~(\ref{etawitha}) is a very significant enhancement over the expected standard-model rate,
\,${\cal B}_{\rm SM}(\eta\to\pi^+\pi^-\mu^+\mu^-)=
\bigl(7.5^{+4.5}_{-2.7}\bigr)\times 10^{-9}$\,~\cite{Faessler:1999de,Borasoy:2007dw},
and may be accessible to DA$\Phi$NE~\cite{Bloise:2007zz}.

\section{Summary and Conclusions}

We have studied several rare decay modes involving a light $CP$-odd Higgs boson in the NMSSM.
In the analysis, for the flavor-changing modes, we have consistently included the two-quark contributions
in which the flavor change occurs in one-loop processes involving the light $A_1^0$ and
the four-quark contributions in which the flavor change occurs via a tree-level standard-model $W$
exchange with the light $A_1^0$ radiated off one of the light quarks.
The interplay between these two contributions was crucial to evade the bounds arising from the nonobservation
of the $A_1^0$ in  \,$K\to\pi\mu^+\mu^-$\, modes in our previous analysis~\cite{He:2005we}.

For the two-quark contributions, we have considered a somewhat general scenario in which the coefficients
$C_{L,R}$ are effectively independent.
We have started with the large-$\tan\beta$ limit where chargino-mediated one-loop diagrams dominate,
but we have also allowed for the possibility of having sizable neutralino- and gluino-mediated one-loop
diagrams.
In this more general scenario, it would also be possible to evade the \,$K\to\pi\mu^+\mu^-$\, bounds
even if the four-quark contributions were absent.

We have evaluated the rare modes  \,$K_L^{}\to\pi\pi A_1^0$\,  which depend on both the two- and
four-quark contributions.
We have found that their rates are significant for most of the allowed parameter space.
These modes are of immediate interest because they can be studied with KTeV data.
It is expected that these studies can help confirm or refute the light $A_1^0$ hypothesis as
a candidate to explain the HyperCP events in  \,$\Sigma^+ \to p \mu^+\mu^-$.\,

We have also studied the modes  \,$\eta\to\pi\pi A_1^0$\,  which depend only on the tree-level
couplings of the $A_1^0$.
Therefore, the predictions for these modes are much less model-dependent and should be of interest
for future experiments at DA$\Phi$NE.
In particular, the $A_1^0$-mediated contribution to  \,$\eta\to\pi^+\pi^-\mu^+\mu^-$\,  can be much
larger than the SM contribution.

Finally, we have revisited the mode  \,$\Omega^-\to\Xi^-A_1^0\to\Xi^-\mu^+\mu^-$\,  to include both 
the two- and four-quark contributions.
We have found that its decay rate could be substantially enhanced with respect to the 
\,$\Omega^-\to\Xi^-\mu^+\mu^-$\, rate in the SM.
This should give additional motivation for conducting experimental studies on the $\Omega^-$ 
in the future.

\begin{acknowledgments}

The work of X.G.H. was supported in part by NSC and NCTS.
The work of G.V. was supported in part by DOE under contract number DE-FG02-01ER41155.
X.G.H. thanks C.-S. Li for useful discussions.
J.T. thanks CTS at NTU for its hospitality during the completion of this work.
G.V. thanks the Cavendish Laboratory at the University of Cambridge and CERN for their hospitality
while this work was completed.

\end{acknowledgments}

\appendix

\section{Chiral Lagrangians for various interactions\label{lags}}

The Lagrangians we have collected here contain not only the baryon- and meson-octet fields,
but also the baryon-decuplet fields.
Since we already derived or used some of the following formulas in Refs.~\cite{He:2005we,He:2006uu},
further details can be found therein.

The chiral realization of  ${\cal L}_{{\cal A}sd}$ in  Eq.~(\ref{LAsd}) can be obtained
employing the operator matching of Ref.~\cite{He:2005we}.
Thus at leading order
\begin{eqnarray}  \label{LA}
{\cal L}_{\cal A}^{} &=&
b_D^{} \left\langle \bar{B}{}^{} \left\{ h_{\cal A}^{}, B^{} \right\} \right\rangle
+ b_F^{} \left\langle \bar{B}{}^{} \left[ h_{\cal A}^{}, B^{} \right] \right\rangle
+ b_0^{} \left\langle h_{\cal A}^{} \right\rangle \left\langle\bar{B}{}^{}B^{}\right\rangle
\,\,+\,\,  \mbox{$\frac{1}{2}$} f^2 B_0^{} \left\langle h_{\cal A}^{} \right\rangle
\nonumber \\ && +\,\,
c\, \bar{T}{}^\alpha h_{\cal A}^{} T_{\alpha}^{}
- c_0^{} \left\langle h_{\cal A}^{} \right\rangle \bar{T}{}^\alpha T_{\alpha}^{}
\,\,+\,\,  {\rm H.c.}    \,\,,
\end{eqnarray}
where  \,$\,f=f_\pi^{}=92.4\rm\,MeV$,\,  \,$B_0^{}=2031$\,MeV,\, and
\begin{eqnarray}
h_{\cal A}^{} \,\,=\,\,
-i \bigl(C_R^{}\,\xi^\dagger h\xi^\dagger+C_L^{}\,\xi h\xi\bigr)\,A_1^0 \,\,.
\end{eqnarray}

To derive amplitudes, we also need the chiral Lagrangian for the strong interactions of
the hadrons~\cite{Gasser:1983yg,bcpt}.
At lowest order in the derivative and $m_s^{}$ expansions, it can be expressed as
\begin{eqnarray}   \label{Ls}
{\cal L}_{\rm s}^{}  &=&
\left\langle \bar{B}^{}\, {i}\gamma^\mu \bigl(
\partial_\mu^{}B+\bigl[{\cal V}_\mu^{},B \bigr] \bigr) \right\rangle
- m_0^{} \left\langle \bar{B}^{} B^{} \right\rangle
%\nonumber \\ && +\,\,
\,+\,
D \left\langle \bar{B}^{} \gamma^\mu\gamma_5^{}
 \left\{ {\cal A}_\mu^{}, B^{} \right\} \right\rangle
+ F \left\langle \bar{B}^{} \gamma^\mu\gamma_5^{}
 \left[ {\cal A}_\mu^{}, B^{} \right] \right\rangle
\nonumber \\ && +\,\,
b_D^{} \left\langle\bar{B}^{}\left\{M_+,B^{}\right\}\right\rangle
+ b_F^{} \left\langle\bar{B}^{}\left[M_+,B^{}\right]\right\rangle
+ b_0^{} \left\langle M_+ \right\rangle \left\langle \bar{B}^{} B^{} \right\rangle
\,+\,
\mbox{$\frac{1}{4}$} f^2 \left\langle
\partial^\mu\Sigma^\dagger\, \partial_\mu^{}\Sigma \right\rangle +
\mbox{$\frac{1}{2}$} f^2 B_0^{} \left\langle M_+ \right\rangle
\nonumber \\ && -\,\,
\bar T^\mu\, {\rm i}\!\not{\!\!\cal D} T_\mu^{} + m_T^{}\; \bar T^\mu T_\mu^{}
+ {\cal C} \left( \bar T^\mu {\cal A}_\mu^{} B + \bar B {\cal A}_\mu^{} T^\mu \right)
\,+\,
c\, \bar T^\mu M_+^{} T_\mu^{} - c_0^{} \left\langle M_+^{}\right\rangle \bar T^\mu T_\mu^{} \,\,,
\end{eqnarray}
where only the relevant terms are displayed and  \,$|{\cal C}|=1.7$.\,

The leading-order Lagrangians relevant to the four-quark interactions involving the $A_1^0$ can
be derived, following the prescription described in Refs.~\cite{Grzadkowski:1992av,He:2006uu},
from  ${\cal L}_{\rm s}^{}$ above and from the mass term in the leading-order Lagrangian
for the  $\,|\Delta I|=\frac{1}{2}\,$ component of the effective Hamiltonian transforming as
$(8_{\rm L}^{},1_{\rm R}^{})$, namely~\cite{Cronin:1967jq}
\begin{eqnarray} \label{Lw}
{\cal L}_{\rm w}^{} &=&
h_D^{} \left\langle \bar B \left\{\xi^\dagger h \xi,\,B \right\} \right\rangle
+ h_F^{} \left\langle \bar B \left[\xi^\dagger h \xi,\,B \right] \right\rangle
\,+\, \gamma_8^{}f^2 \left\langle h\,\partial_\mu^{}\Sigma\, \partial^\mu \Sigma^\dagger \right\rangle
+ 2\tilde\gamma_8^{} f^2 B_0^{} \left\langle h \xi M_+^{} \xi^\dagger \right\rangle
\nonumber \\ &&
+\,\,
h_C^{}\, \bar T^\mu \xi^\dagger h \xi T_\mu^{}
\,\,+\,\,  {\rm H.c.}  \,\,,
\end{eqnarray}
where  \,$\gamma_8^{}=-7.8\times10^{-8}$.\,
Thus we have
\begin{eqnarray}   \label{LsP}
{\cal L}_{\rm s}^{\cal A}  &=&
\Bigl( b_D^{}\,\bigl\langle \bar{B}^{} \bigl\{ \tilde{M}_-^{}, B^{} \bigr\} \bigr\rangle
+ b_F^{}\, \bigl\langle \bar{B}^{} \bigl[ \tilde{M}_-^{}, B^{} \bigr] \bigr\rangle
+ b_0^{}\, \bigl\langle \tilde{M}_-^{} \bigr\rangle \bigl\langle \bar{B}^{} B^{} \bigr\rangle
\,+\, \mbox{$\frac{1}{2}$} f^2 B_0^{}\, \bigl\langle\tilde{M}_-^{}\bigr\rangle
\nonumber \\ && \;\;+\,\,
c\, \bar T^\mu \tilde{M}_-^{} T_\mu^{}
- c_0^{}\, \bigl\langle \tilde{M}_-^{} \bigr\rangle\, \bar T^\mu T_\mu^{} \Bigr)
\frac{{i}A_1^0}{v}  \,\,,
\end{eqnarray}
\begin{eqnarray} \label{LwP}
{\cal L}_{\rm w}^{\cal A}  \,\,=\,\,
2\tilde{\gamma}_8^{}\, f^2 B_0^{}\, \bigl\langle h \xi \tilde{M}_-^{}\xi^\dagger
\bigr\rangle \frac{i A_1^0}{v}
\,\,+\,\,  {\rm H.c.}   \,\,,
\end{eqnarray}
where  \,$\tilde{M}_-^{}=\xi^\dagger\tilde{M}\xi^\dagger-\xi\tilde{M}^\dagger\xi$,\,  with
\,$\tilde{M}={\rm diag}\bigl(0,\,l_d^{}{}\hat{m},\,l_d^{}{}m_s^{}\bigr)\,$  at large $\tan\beta$.
We include the SU(3) singlet $\eta_1^{}$ in  ${\cal L}_{\rm s,w}^{({\cal A})}$
by replacing  $\Sigma$ with  \,$\Sigma\,\exp\bigl(i\sqrt{2/3}\,\eta_1^{}/f\bigr)$\,
and adding the anomaly-generated term
\begin{eqnarray} \label{LeP}
{\cal L}_{\eta_1^{}{\cal A}}^{}  \,\,=\,\,
-\frac{\tilde m_0^2}{2} \Biggl(\eta_1^{}+\frac{f\,A_1^0\,l_d^{}}{\sqrt6\,v}\Biggr)^{\!2} \,\,,
\end{eqnarray}
which modifies the $\eta_1^{}$-$A_1^0$ mixing generated by  ${\cal L}_{\rm s}^{\cal A}$.
The physical $\eta$ and $\eta'$ fields are related to $\eta_1^{}$ and the SU(3) octet $\eta_8^{}$ by
\,$\eta=\eta_8^{}\,c_\theta^{}-\eta_1^{}\,s_\theta^{}$\,  and
\,$\eta'=\eta_8^{}\,s_\theta^{}+\eta_1^{}\,c_\theta^{}$,\,  where  \,$c_\theta^{}=\cos\theta$\,
and  \,$s_\theta^{}=\sin\theta$.\,
As in Ref.~\cite{He:2006uu}, we adopt $\,\tilde{m}_0^{}=819$\,MeV\, and $\,\theta=-19.7^\circ$.\,

\section{Four-quark contributions to \,$\bm{\bar K^0\to\pi\pi A_1^0}$\label{Mi}}

In the case of  \,$\bar K^0\to\pi^+\pi^-A_1^0$,\,  for \,$l_u^{}=0$\,  the eight diagrams in
Fig.~\ref{4qdiag} yield, respectively,
% 1
\begin{eqnarray}
{\cal M}_1^{+-}  \,\,=\,\,  \frac{\sqrt8\, m_K^2}{3 f v}\, \tilde\gamma_8^{}\,l_d^{}  \,\,,
\end{eqnarray}
% 2
\begin{eqnarray}
{\cal M}_2^{+-}  \,\,=\,\,  \frac{\sqrt2\, m_K^2}{3 f v}\,\,
\frac{3 m_{{\cal A}\pi^+}^2-3 m_\pi^2-2 m_K^2-m_{\cal A}^2}{m_K^2-m_{\cal A}^2}\,\,
\tilde\gamma_8^{}\,l_d^{}  \,\,,
\end{eqnarray}
% 3
\begin{eqnarray}
{\cal M}_3^{+-}  \,\,=\,\,  \frac{\sqrt8\, m_K^2}{3 f v}\,\,
\frac{\gamma_8^{}\, m_\pi^2 - \tilde\gamma_8^{}\, m_K^2}{m_K^2-m_\pi^2}\,\, l_d^{}  \,\,,
\end{eqnarray}
% 4
\begin{eqnarray}
{\cal M}_4^{+-}  &=&
\frac{\sqrt2}{36 f v} \Bigl[
9b_\pi^{}\, \bigl(m_{{\cal A}\pi^-}^2-m_{\pi^+\pi^-}^2\bigr)
\,-\,
\bigl( b_\eta^{} c_\theta^{}+b_{\eta'}^{} s_\theta^{} \bigr)
\bigl( 5 m_K^2+4 m_\pi^2+3 m_{\cal A}^2-9 m_{{\cal A}\pi^+}^2 \bigr)
\nonumber \\ && \hspace*{7ex}
-\,\,
\sqrt8\, \bigl(b_\eta^{} s_\theta^{}-b_{\eta'}^{} c_\theta^{}\bigr) \bigl(2 m_K^2+m_\pi^2\bigr)
\Bigr] \frac{\gamma_8^{}\, m_\pi^2-\tilde\gamma_8^{}\, m_K^2}{m_K^2-m_\pi^2} \, l_d^{} \,\,,
\end{eqnarray}
% 5
\begin{eqnarray}
{\cal M}_5^{+-}  &=&
\frac{\sqrt2\, m_K^2}{6 f v} \Biggl[ \frac{-m_\pi^2}{m_K^2-m_\pi^2}
\,+\,
\frac{m_\pi^2}{m_K^2-m_\eta^2}
\bigl(c_\theta^{}-\sqrt2\,s_\theta^{}\bigr)\bigl(c_\theta^{}+\sqrt8\,s_\theta^{}\bigr)
\nonumber \\ && \hspace*{9ex}
+\,\,
\frac{m_\pi^2}{m_K^2-m_{\eta'}^2}
\bigl(\sqrt2\, c_\theta^{}+s_\theta^{}\bigr)\bigl(s_\theta^{}-\sqrt8\, c_\theta^{}\bigr)
\Biggr] \bigl(\gamma_8^{}-\tilde\gamma_8^{}\bigr)\, l_d^{}  \,\,,
\end{eqnarray}
% 6
\begin{eqnarray}
{\cal M}_6^{+-}  &=&
\frac{\sqrt2}{36 f v} \Bigl[ 3 b_\pi^{}\, \bigl(2m_K^2+2m_{\cal A}^2-3m_{\pi^+\pi^-}^2\bigr)
\,+\,
\bigl( b_\eta^{} c_\theta^{}+b_{\eta'}^{} s_\theta^{} \bigr)
\bigl(2 m_{\cal A}^2-6 m_K^2-3 m_{\pi^+\pi^-}^2+6 m_{{\cal A}\pi^-}^2\bigr)
\nonumber \\ && \hspace*{7ex}
-\,\,
\sqrt8\, \bigl( b_\eta^{} s_\theta^{}-b_{\eta'}^{} c_\theta^{} \bigr)
\bigl(3 m_\pi^2+m_{\cal A}^2-3 m_{{\cal A}\pi^+}^2\bigr)
\Bigr] \gamma_8^{}\, l_d^{}
\nonumber \\ &&
-\,\,
\frac{\sqrt2\, m_K^2}{36 f v} \Bigl[
3 b_\pi^{} \,-\, b_\eta^{}\, \bigl(c_\theta^{}-4\sqrt2\, s_\theta^{}\bigr)
\,-\, b_{\eta'}^{}\,\bigl(4\sqrt2\,c_\theta^{}+s_\theta^{}\bigr)\Bigr] \tilde\gamma_8^{}\,l_d^{}\,\,,
\end{eqnarray}
% 7
\begin{eqnarray}
{\cal M}_7^{+-}  &=&
\frac{\sqrt2}{36 f v} \Bigl[
3 b_\pi^{} \,+\, b_\eta^{}\, \bigl(c_\theta^{}+\sqrt8\, s_\theta^{}\bigr)
\,-\, b_{\eta'}^{}\, \bigl(\sqrt8\, c_\theta^{}-s_\theta^{}\bigr) \Bigr]
\bigl(\gamma_8^{}\, m_{\cal A}^2-\tilde\gamma_8^{}\, m_K^2\bigr) l_d^{}
\nonumber \\ &&
\times\,\,
\frac{3 m_{{\cal A}\pi^+}^2-2 m_K^2-3 m_\pi^2-m_{\cal A}^2}{m_K^2-m_{\cal A}^2} \,\,,
\end{eqnarray}
% 8
\begin{eqnarray}
{\cal M}_8^{+-}  &=&
\frac{\sqrt2\, m_K^2}{18 f v} \Biggl\{
3 b_\pi^{}\, \frac{3 m_{\pi^+\pi^-}^2-m_K^2-m_\pi^2-m_{\cal A}^2}{m_K^2-m_\pi^2}
\nonumber \\ && \hspace*{5ex}
+\,\,
\Bigl[ b_\eta^{}\, \bigl(1-\sqrt8\, c_\theta^{} s_\theta^{}+s_\theta^2\bigr)
      + b_{\eta'}^{}\, \bigl(\sqrt2-c_\theta^{} s_\theta^{}-\sqrt8\,s_\theta^2\bigr) \Bigr]
\frac{\bigl(c_\theta^{}+\sqrt8\, s_\theta^{}\bigr)m_\pi^2}{m_K^2-m_\eta^2}
\nonumber \\ && \hspace*{5ex}
-\,\,
\Bigl[ b_\eta^{}\, \bigl(\sqrt2-c_\theta^{} s_\theta^{}-\sqrt8\,s_\theta^2\bigr)
      + b_{\eta'}^{}\, \bigl(2+\sqrt8\,c_\theta^{} s_\theta^{}-s_\theta^2\bigr) \Bigr]
\frac{\bigl(\sqrt8\,c_\theta^{}-s_\theta^{}\bigr) m_\pi^2}{m_K^2-m_{\eta'}^2}
\Biggr\} \bigl(\gamma_8^{}-\tilde\gamma_8^{}\bigr) l_d^{} \,\,,
\nonumber \\
\end{eqnarray}
where  \,$m_{XY}^2=(p_X^{}+p_Y^{})^2$,\,
\begin{eqnarray}
b_\pi^{} \,\,=\,\, \frac{m_\pi^2}{m_\pi^2-m_{\cal A}^2}  \,\,,
\end{eqnarray}
\begin{eqnarray}
b_\eta^{} \,\,=\,\,
\frac{\bigl(4m_K^2-3m_\pi^2\bigr)c_\theta^{}+\sqrt2\,\bigl(2m_K^2-\tilde m_0^2\bigr)s_\theta^{}}
     {m_\eta^2-m_{\cal A}^2} \,\,,
\end{eqnarray}
\begin{eqnarray}
b_{\eta'}^{} \,\,=\,\,
\frac{\bigl(4m_K^2-3m_\pi^2\bigr)s_\theta^{}-\sqrt2\,\bigl(2m_K^2-\tilde m_0^2\bigr)c_\theta^{}}
     {m_{\eta'}^2-m_{\cal A}^2}  \,\,.
\end{eqnarray}
In the case of  \,$\bar K^0\to\pi^0\pi^0A_1^0$,\,  the diagrams in Fig.~\ref{4qdiag} yield,
for  \,$l_u^{}=0$,\,
% 1
\begin{eqnarray}
{\cal M}_1^{00}  \,\,=\,\,  \frac{\sqrt8\, m_K^2}{3 f v}\, \tilde\gamma_8^{}\,l_d^{}  \,\,,
\end{eqnarray}
% 2
\begin{eqnarray}
{\cal M}_2^{00}  \,\,=\,\,  \frac{\sqrt2\, m_K^2}{6 f v}\,\,
\frac{m_{\cal A}^2-m_K^2-3 m_{\pi^0\pi^0}^2}{m_K^2-m_{\cal A}^2}\,\, \tilde\gamma_8^{}\,l_d^{}  \,\,,
\end{eqnarray}
% 3
\begin{eqnarray}
{\cal M}_3^{00}  \,\,=\,\,  \frac{\sqrt2\, (2 m_K^2+m_\pi^2)}{3 f v}\,\,
\frac{\gamma_8^{}\, m_\pi^2 - \tilde\gamma_8^{}\, m_K^2}{m_K^2-m_\pi^2}\,\, l_d^{}  \,\,,
\end{eqnarray}
% 4
\begin{eqnarray}
{\cal M}_4^{00}  &=&
\frac{\sqrt2}{72 f v} \Bigl[
3 b_\pi^{}\, \bigl(3 m_{\pi^0\pi^0}^2-5 m_K^2-6 m_\pi^2-m_{\cal A}^2 \bigr)
\,-\,
\bigl( b_\eta^{} c_\theta^{}+b_{\eta'}^{} s_\theta^{} \bigr)
\bigl( m_K^2-10 m_\pi^2-3 m_{\cal A}^2+9 m_{\pi^0\pi^0}^2 \bigr)
\nonumber \\ && \hspace*{7ex}
-\,\,
4\sqrt2\, \bigl( b_\eta^{}\, s_\theta^{}-b_{\eta'}^{}\, c_\theta^{} \bigr) \bigl( 2 m_K^2+m_\pi^2 \bigr)
\Bigr] \frac{\gamma_8^{}\, m_\pi^2 \,-\, \tilde\gamma_8^{}\, m_K^2}{m_K^2-m_\pi^2}\,\, l_d^{} \,\,,
\end{eqnarray}
% 5
\begin{eqnarray}
{\cal M}_5^{00}  &=&
\frac{\sqrt2\, m_K^2}{6 f v} \Biggl[ \frac{-3 m_\pi^2}{m_K^2-m_\pi^2}
\,+\,
\frac{m_\pi^2}{m_K^2-m_\eta^2}
\bigl(c_\theta^{}-\sqrt2\,s_\theta^{}\bigr)\bigl(c_\theta^{}+\sqrt8\,s_\theta^{}\bigr)
\nonumber \\ && \hspace*{9ex}
+\,\,
\frac{m_\pi^2}{m_K^2-m_{\eta'}^2}
\bigl(\sqrt2\, c_\theta^{}+s_\theta^{}\bigr)\bigl(s_\theta^{}-\sqrt8\, c_\theta^{}\bigr)
\Biggr] \bigl(\gamma_8^{}-\tilde\gamma_8^{}\bigr) l_d^{}  \,\,,
\end{eqnarray}
% 6
\begin{eqnarray}
{\cal M}_6^{00}  &=&
\frac{\sqrt2}{36 f v} \Bigl[ 3 b_\pi^{}\, \bigl(3 m_K^2-2m_\pi^2-m_{\cal A}^2\bigr)
\,-\,
\bigl( b_\eta^{} c_\theta^{}+b_{\eta'}^{} s_\theta^{} \bigr)
\bigl(3 m_K^2-6 m_\pi^2-5 m_{\cal A}^2+6 m_{\pi^0\pi^0}^2\bigr)
\nonumber \\ && \hspace*{3em}
+\,\,
\sqrt2\, \bigl( b_\eta^{} s_\theta^{}-b_{\eta'}^{} c_\theta^{} \bigr)
\bigl(3 m_K^2+m_{\cal A}^2-3 m_{\pi^0\pi^0}^2\bigr)
\Bigr] \gamma_8^{}\, l_d^{}
\nonumber \\ && \!\!\! \!
-\,\,
\frac{\sqrt2\, m_K^2}{36 f v} \Bigl[
9 b_\pi^{} \,-\, b_\eta^{}\, \bigl(c_\theta^{}-4\sqrt2\, s_\theta^{}\bigr)
\,-\, b_{\eta'}^{}\, \bigl(4\sqrt2\, c_\theta^{}+s_\theta^{}\bigr) \Bigr] \tilde\gamma_8^{}\, l_d^{}
\end{eqnarray}
% 7
\begin{eqnarray}
{\cal M}_7^{00}  &=&
\frac{\sqrt2}{72 f v} \Bigl[
3 b_\pi^{} \,+\, b_\eta^{}\, \bigl(c_\theta^{}+\sqrt8\, s_\theta^{}\bigr)
\,-\, b_{\eta'}^{}\, \bigl(\sqrt8\, c_\theta^{}-s_\theta^{}\bigr) \Bigr]
\bigl(\gamma_8^{}\, m_{\cal A}^2-\tilde\gamma_8^{}\, m_K^2\bigr) l_d^{}\,\,
\frac{m_{\cal A}^2-m_K^2-3 m_{\pi^0\pi^0}^2}{m_K^2-m_{\cal A}^2} \,\,,
\nonumber \\
\end{eqnarray}
% 8
\begin{eqnarray}
{\cal M}_8^{00}  &=&
\frac{\sqrt2\, m_K^2}{18 f v} \Biggl\{ \frac{9 b_\pi^{}\, m_\pi^2}{m_K^2-m_\pi^2}
\,+\,
\Bigl[ b_\eta^{}\, \bigl(1-\sqrt8\, c_\theta^{} s_\theta^{}+s_\theta^2 \bigr)
      + b_{\eta'}^{}\, \bigl(\sqrt2-c_\theta^{} s_\theta^{}-\sqrt8\, s_\theta^2\bigr) \Bigr]
\frac{\bigl(c_\theta^{}+\sqrt8\, s_\theta^{}\bigr) m_\pi^2}{m_K^2-m_\eta^2}
\nonumber \\ && \hspace*{5ex}
-\,\,
\Bigl[ b_\eta^{}\, \bigl(\sqrt2-c_\theta^{} s_\theta^{}-\sqrt8\, s_\theta^2\bigr)
      + b_{\eta'}^{}\, \bigl(2+\sqrt8\,c_\theta^{}\,s_\theta^{}-s_\theta^2 \bigr) \Bigr]
\frac{\bigl(\sqrt8\,c_\theta^{}-s_\theta^{}\bigr) m_\pi^2}{m_K^2-m_{\eta'}^2}
\Biggr\} \bigl(\gamma_8^{}-\tilde\gamma_8^{}\bigr) l_d^{} \,\,,
\nonumber \\
\end{eqnarray}
In the expressions for ${\cal M}_i^{+-}$ or ${\cal M}_i^{00}$ above, we have kept the terms
proportional to $\tilde\gamma_8^{}$ in order to check our algebra.
As explained in Ref.~\cite{Grzadkowski:1992av}, the $\tilde{\gamma}_8^{}$ terms in
${\cal L}_{\rm w}^{({\cal A})}$ can be rotated away for kaon decay, and we have verified that
the $\tilde{\gamma}_8^{}$ terms cancel accordingly in the sum of the contributions.
In our numerical evaluation, $\tilde{\gamma}_8^{}$ is thus set to zero.

\end{document}